\documentclass[12pt]{article}
\usepackage{amssymb,amsmath,epsfig}
\allowdisplaybreaks
\begin{document}
\title{\bf Gravitational Decoupled Anisotropic Solutions in $f(\mathcal{G})$
Gravity}
\author{M. Sharif \thanks{msharif.math@pu.edu.pk} and Saadia Saba
\thanks{saadia.saba86@gmail.com}\\
Department of Mathematics, University of the Punjab,\\
Quaid-e-Azam Campus, Lahore-54590, Pakistan.}
\date{}

\maketitle

\begin{abstract}
In this paper, we investigate anisotropic static spherically
symmetric solutions in the framework of $f(\mathcal{G})$ gravity
through gravitational decoupling approach. For this purpose, we
consider Krori and Barua (known solution) isotropic interior
solution for static spherically symmetric self-gravitating system
and extend it to two types of anisotropic solutions. We examine the
physical viability of our models through energy conditions, squared
speed of sound and anisotropy parameter. It is found that the first
solution is physically viable as it fulfills the energy bounds as
well as stability criteria while the second solution satisfies all
energy bounds but is unstable at the core of the compact star.
\end{abstract}

\section{Introduction}

Modified theories of gravity have secured extensive recognition
after the innovative cosmological aspects of the expanding universe.
This intriguing approach is considered as the most promising and
optimistic to unveil the hidden characteristics of cosmos. Nojiri
and Odintsov \cite{2} introduced modified Gauss-Bonnet gravity (or
$f(\mathcal{G})$ gravity) by including higher order correction terms
through Gauss-Bonnet (GB) invariant. The motivation behind this
theory comes from string theory at low energy scales which is
expected to analyze effectively the late-time cosmic transitions.
The GB invariant is a four-dimensional topological term which is a
combination of the Ricci scalar ($R$), Ricci
($R_{\alpha\beta}$) and Riemann tensors ($R_{\alpha\beta\mu\nu}$),
given by
$\mathcal{G}=R^{2}-4R_{\alpha\beta}R^{\alpha\beta}+R_{\alpha\beta\mu\nu}
R^{\alpha\beta\mu\nu}$. This second Lovelock scalar trivially
contributes when included in matter Lagrangian but excludes spin-2
ghost instability \cite{3}. Bamba et al. \cite{5} explored modified
$f(\mathcal{G})$ as well as $f(R,\mathcal{G})$ models with some
emergent ingredients of finite-time future singularities and
investigated higher-order curvature corrections to cure these
singularities.

The formulation of appropriate spherically symmetric interior
solution of self-gravitating system has always been problematic due
to the presence of nonlinearity in the field equations. A plenty of
work has been done in literature to tackle this issue. Mak and Harko
\cite{6} obtained exact anisotropic solution of the field equations
and found the positively finite behavior of density and pressure
supporting the core of stellar objects. Gleiser and Dev \cite{7}
investigated algorithm of anisotropic self-gravitating system with
compactness $\frac{M}{R}=\frac{4}{9}$ and obtained stable model for
small measures of adiabatic index. Sharma and Maharaj \cite{8}
explored some anisotropic spherical exact solutions in the framework
of linear combination of equation of state defining compactness of
stellar objects. Kalam et al. \cite{9} found compact models in the
context of anisotropic regime using Karori and Barua metric. Bhar et
al. \cite{10} discussed the possibilities of existence of compact
objects in higher dimensions. Maurya et al. \cite{11} studied
anisotropic solutions of compact stars in the presence of charge
distribution.

The existence of exact interior solutions of self-gravitating
systems in the presence of anisotropy have been carried out in a
number of ways. In this regard, the minimal gravitational decoupling
(MGD) approach appeared as significantly deterministic in finding
the physically viable solutions for spherically symmetric stellar
configuration. Ovalle \cite{12} proposed this technique to extract
some exact solutions for compact stellar objects in the framework of
braneworld. The MGD approach is not a novel idea, indeed some
ingredients make it specifically more attractive for searching new
spherically symmetric solutions of the Einstein field equations. The
main and foremost feature of this technique is that a simple
solution can be extended into more complex domains. This technique
could be started with a simple source ($T^{(m)}_{\mu\nu}$) in which
another gravitational source ($T^{\ast}_{\mu\nu}$) can be added
through a coupling constant $\alpha$, i.e.,
$T^{(m)}_{\mu\nu}\rightarrow\tilde{T}^{(tot)}_{\mu\nu}=T^{(m)}_{\mu\nu}+\alpha
T^{\ast}_{\mu\nu}$ such that the spherical symmetry remains
preserved. The reverse of this technique also works through
de-coupling of gravitationally sources. In order to find the
solution of highly non-linear field equations with complex
spherically symmetric gravitational sources, we split the source in
simple components and find the solution for each of them. This leads
to as many solutions as the number of components whose combination
will yield the solution of the field equation corresponding to the
original energy-momentum tensor. This technique provides a
breakthrough in the search of anisotropic solutions extended from
isotropic ones.

In this context, Ovalle and Linares \cite{13} formulated
an exact solution of the field equations for spherically symmetric
isotropic compact distribution and concluded that their results
present the braneworld form of Tolman-IV solution. Casadio et al.
\cite{14} developed some exterior solutions for spherically
symmetric self-gravitating system using gravitational decoupling
technique and found naked singularity at Schwarzschild radius.
Ovalle \cite{15} decoupled gravitational source to obtain
anisotropic solutions from spherically symmetric isotropic
solutions. Ovalle et al. \cite{16} extended the isotropic solution
by inclusion of anisotropy using MGD approach for static regime of
stellar objects. Sharif and Sadiq \cite{17} explored charged
anisotropic spherical solution through this approach and also
examined the viability conditions, stability criteria through
squared speed of sound.

Compact stars being the relativistic massive objects (small size and
extremely massive structure) possess very strong gravitational force
which can be studied in modified theories of gravity. The curiosity
to know more about compact stars brings in many researchers on the
platform of modified theories of gravity. Zubair and Abbas \cite{18}
investigated the possibilities of formulation of compact star in
$f(R)$ gravity using Karori and Barua solution. Abbas and his
collaborators \cite{19} analyzed the anisotropic compact star
solution in $f(T)$ gravity and examined the surface redshift,
stability as well as regularity conditions. Abbas et al. \cite{20}
analyzed the anisotropic compact star in $f(\mathcal{G})$
gravity and examined physical behavior of star with observational
data. Sharif and Fatima \cite{21} explored static spherically
symmetric solutions in $f(\mathcal{G})$ gravity for both
isotropic and anisotropic matter distributions.

In this paper, we explore anisotropic spherically symmetric
solutions using MGD approach. The paper is organized in the
following format. In the next section, we discuss some basic
terminologies of $f(\mathcal{G})$ gravity and corresponding field equations
for multiple sources. Section \textbf{3} is devoted to MGD approach
and corresponding junction conditions. In section \textbf{4}, we
find exact anisotropic solutions using some constraints and check
their physical behavior. Finally, we conclude our results in the
last section.

\section{Fluid Configuration and Field Equations for Multiple Sources}

The standard field equation for $f(\mathcal{G})$ gravity are
\cite{22}
\begin{equation}\label{1}
R_{\rho\gamma}-\frac{1}{2} R g_{\rho\gamma}=\kappa
T_{\rho\gamma}^{(tot)},
\end{equation}
where $\kappa$ is coupling constant with
\begin{equation}\label{1a}
T_{\rho\gamma}^{(tot)}=T_{\rho\gamma}^{(m)}+T_{\rho\gamma}^{(\mathcal{G})}+\alpha\Theta_{\rho\gamma}.
\end{equation}
The energy-momentum tensor for perfect fluid configuration
containing four-velocity field, density and pressure is
\begin{equation}\label{1b}
T_{\rho\gamma}^{(m)}=(\rho+P)U_{\rho}U_{\gamma}+Pg_{\rho\gamma},
\end{equation}
and
\begin{eqnarray}\nonumber
T_{\rho\gamma}^{(\mathcal{G})}&=&
\frac{1}{2\kappa}g_{\rho\gamma}f(\mathcal{G})+\frac{1}{\kappa}[(4R_{\gamma\mu}
R^{\mu}_{\rho}-2RR_{\rho\gamma}-2R_{\rho\mu\eta\nu}R^{\mu\eta\nu}_{\gamma}
-4R_{\rho\mu\eta\gamma}R^{\mu\eta})f_{\mathcal{G}}(\mathcal{G})\\\nonumber&+&
(4R_{\rho\gamma}-2Rg_{\rho\gamma})\nabla^{2}
f_{\mathcal{G}}(\mathcal{G})+
2R\nabla_{\rho}\nabla_{\gamma}f_{\mathcal{G}}(\mathcal{G})\\\nonumber&-&
4R_{\rho}^{\mu}\nabla_{\gamma}\nabla_{\mu}f_{\mathcal{G}}(\mathcal{G})-
4R_{\gamma}^{\mu}\nabla_{\rho}\nabla_{\mu}f_{\mathcal{G}}(\mathcal{G})\\\label{2}&+&
4g_{\rho\gamma}R^{\mu\eta}\nabla_{\mu}\nabla_{\eta}f_{\mathcal{G}}(\mathcal{G})-
4R_{\rho\mu\gamma\eta}\nabla^{\mu}\nabla^{\eta}f_{\mathcal{G}}(\mathcal{G})],
\end{eqnarray}
where $\nabla^{2}=\nabla_{\mu}\nabla^{\mu}$ ($\nabla_{\mu}$ denotes
covariant derivative) is d'Alembert operator and
$f_{\mathcal{G}}(\mathcal{G})$ represents derivative of generic
function with respect to $\mathcal{G}$. The term
$\Theta_{\rho\gamma}$ describes an additional source coupling the
gravity through constant $\alpha$ \cite{23} which may incorporate
new fields (like scalar, vector or tensor fields) and will produce
anisotropy in self-gravitating systems.

The line element of static spherically symmetric spacetime reads
\begin{equation}\label{3}
ds^{2}=-e^{\eta(r)}dt^{2}+e^{\psi(r)}dr^{2}+r^{2}(d\theta^{2}+\sin^{2}\theta
d\phi^{2}),
\end{equation}
where $\eta=\eta(r)$, $\psi=\psi(r)$ denote the function of areal
radius $r$ ranging from core to the surface of star while
$U^{\mu}=e^{-\frac{\eta}{2}}\delta^{\mu}_{0}$ for $0<r<R$. The field
equations (\ref{1}) with (\ref{3}) yield
\begin{eqnarray}\label{4}
\kappa (\rho-T_{0}^{0(\mathcal{G})}-\alpha
\Theta_{0}^{0})&=&\frac{1}{r^{2}}+(\frac{\psi'}{r}-\frac{1}{r^{2}})e^{-\psi(r)},\\\label{5}
\kappa (P+T_{1}^{1(\mathcal{G})}+\alpha
\Theta_{1}^{1})&=&-\frac{1}{r^{2}}+(\frac{\eta'}{r}+\frac{1}{r^{2}})e^{-\psi(r)},\\\label{6}
\kappa (P+T_{2}^{2(\mathcal{G})}+\alpha
\Theta_{2}^{2})&=&(\frac{\eta'}{2r}+\frac{\eta''}{2}+\frac{\eta'^{2}}{4}-\frac{\eta'\psi'}{4}
-\frac{\psi'}{2r})e^{-\psi(r)},
\end{eqnarray}
where $T_{0}^{0(\mathcal{G})}$, $T_{1}^{1(\mathcal{G})}$ and
$T_{2}^{2(\mathcal{G})}$ are given in appendix \textbf{A}. The
expression for GB invariant takes the form
\begin{eqnarray}\label{6ai}
\mathcal{G}=\frac{2e^{-2\psi}}{r^{2}}[(\eta'^{2}+2\eta'')(1-e^{\psi})+\eta'\psi'(e^{\psi})],
\end{eqnarray}
where prime denotes the derivative with respect to $r$. The
corresponding conservation equation reads
\begin{eqnarray}\nonumber
\frac{dP}{dr}&+&\frac{dT_{1}^{1(\mathcal{G})}}{dr}+\alpha\frac{d\Theta_{1}^{1}}{dr}+
\frac{\eta'}{2}(\rho+P+T_{1}^{1(\mathcal{G})}-T_{0}^{0(\mathcal{G})})+\frac{2}{r}(T_{1}^{1(\mathcal{G})}-T_{2}^{2(\mathcal{G})})
\\\label{6a}&+&\frac{\alpha\eta'}{2}(\Theta_{1}^{1}-\Theta_{0}^{0})+\frac{2\alpha}{r}(\Theta_{1}^{1}-\Theta_{2}^{2})=0.
\end{eqnarray}
It is found that the system of non-linear differential equations
((\ref{4})-(\ref{6a})) consists of seven unknown functions ($\psi$,
$\eta$, $\rho$, $P$, $\Theta_{0}^{0}$, $\Theta_{1}^{1}$,
$\Theta_{2}^{2}$). We adopt systematic approach of Ovalle \cite{16}
to determine these unknowns. For the system ((\ref{4})-(\ref{6a})),
the matter contents (effective density, effective isotropic pressure
and effective tangential pressure) can be identified as
\begin{eqnarray}\label{7}
\bar{\rho}=\rho-\alpha\Theta_{0}^{0},\quad
\bar{P}_{r}=P+\alpha\Theta_{1}^{1},\quad
\bar{P}_{t}=P+\alpha\Theta_{2}^{2}.
\end{eqnarray}
This clearly shows that the source $\Theta_{\rho\gamma}$ can, in
general, bring in anisotropy
$\bar{\Delta}=\bar{P}_{t}-\bar{P}_{r}=\alpha(\Theta_{2}^{2}-\Theta_{1}^{1})$
into the inner of stellar distribution.

\section{Gravitational Decoupling by MGD Approach}

In this section, we use MGD approach to find solution of the system
((\ref{4})-(\ref{6a})) by transforming the field equations such that
the source $\Theta_{\rho\gamma}$ takes the form of effective
equations which might incorporate anisotropy. Let us consider the
perfect fluid solution ($\psi$, $\eta$, $\rho$, $P$) with $\alpha=0$
using line element
\begin{equation}\label{8}
ds^{2}=-e^{\chi(r)}dt^{2}+\frac{dr^{2}}{\xi(r)}+r^{2}(d\theta^{2}+\sin^{2}\theta
d\phi^{2}),
\end{equation}
where $\xi(r)=1-\frac{2m}{r}$ contains the Misner-Sharp mass ``$m$"
for the fluid configuration. We take the effects of source
$\Theta_{\rho\gamma}$ in isotropic model by encoding the geometrical
deformation undertaken by perfect fluid metric (\ref{8}) as
\cite{16}
\begin{equation}\label{9}
\chi\rightarrow\eta=\chi+\alpha f, \quad \xi\rightarrow
e^{-\psi}=\xi+\alpha h,
\end{equation}
where $f$ and $h$ are geometrical deformations offered to temporal
and radial metric ingredients. The possibly minimal geometric
deformation among aforementioned deformations is
\begin{equation}\label{10}
f\rightarrow0, \quad h\rightarrow h^{\ast},
\end{equation}
where the radial metric component endures deformation while the
temporal component remains the same. Hence the minimal geometric
deformation equation (\ref{10}) turns out to be
\begin{equation}\label{11}
\chi\rightarrow\eta=\chi, \quad \xi\rightarrow e^{-\psi}=\xi+\alpha
h^{\ast},
\end{equation}
where $h^{\ast}$ is the deformation function associated to radial
metric component. Using Eq.(\ref{11}), the system
((\ref{4})-(\ref{6a})) splits up into two sets.

The first set gives
\begin{eqnarray}\label{12}
\kappa
(\rho-T_{0}^{0(\mathcal{G})})&=&\frac{1}{r^{2}}-(\frac{\xi'}{r}+\frac{\xi}{r^{2}}),\\\label{13}
\kappa(P+T_{1}^{1(\mathcal{G})})&=&-\frac{1}{r^{2}}+(\frac{\chi'}{r}+\frac{1}{r^{2}})\xi(r),\\\label{14}
\kappa(P+T_{2}^{2(\mathcal{G})})&=&(\frac{\chi'}{2r}+\frac{\chi''}{2}+\frac{\chi'^{2}}{4})\xi(r)
+(\frac{\chi'}{4}+\frac{1}{2r})\xi'(r),
\end{eqnarray}
and the second one containing the source is
\begin{eqnarray}\label{15}
\kappa\Theta_{0}^{0}&=&\frac{h^{\ast'}}{r}+\frac{h^{\ast}}{r^{2}},\\\label{16}
\kappa\Theta_{1}^{1}&=&h^{\ast}(\frac{\chi'}{r}+\frac{1}{r^{2}}),\\\label{17}
\kappa\Theta_{2}^{2}&=&(\frac{\chi'}{2r}+\frac{\chi''}{2}+\frac{\chi'^{2}}{4})\xi(r)
+(\frac{\chi'}{4}+\frac{1}{2r})\xi'(r),
\end{eqnarray}
The above system (\ref{15})-(\ref{17}) looks similar to the
spherically symmetric field equations for anisotropic fluid
configuration with source $\Theta_{\rho\gamma}$
($\rho=-\Theta_{0}^{0},~ P_{r}=\Theta_{1}^{1},~
P_{t}=\Theta_{2}^{2}$) corresponding to the metric
\begin{equation}\label{18}
ds^{2}=-e^{\chi(r)}dt^{2}+\frac{dr^{2}}{h^{\ast}(r)}+r^{2}(d\theta^{2}+\sin^{2}\theta
d\phi^{2}).
\end{equation}
However, the right-hand side of system (\ref{15})-(\ref{17})
deviates from the anisotropic solution by term $\frac{1}{r^{2}}$
which constitutes the effective matter components as
\begin{eqnarray}\label{19}
\bar{\rho}=\Theta_{0}^{0\ast}=\Theta_{0}^{0}-\frac{1}{\kappa
r^{2}},\quad
\bar{P}_{r}=\Theta_{1}^{1\ast}=\Theta_{1}^{1}-\frac{1}{\kappa
r^{2}},\quad \bar{P}_{t}=\Theta_{2}^{2\ast}=\Theta_{2}^{2}.
\end{eqnarray}

The junction conditions provide smooth matching of interior and
exterior geometries at the surface of the stellar object to
investigate some significant features of their evolution. For
instance, the interior spacetime geometry of stellar distribution is
obtained through MGD as
\begin{equation}\label{20}
ds^{2}=-e^{\eta_{-}(r)}dt^{2}+(1-\frac{2\bar{m}}{r})^{-1}dr^{2}+r^{2}(d\theta^{2}+\sin^{2}\theta
d\phi^{2}),
\end{equation}
where the interior mass is $\bar{m}=m(r)-\frac{r}{2}\alpha
h^{\ast}$. Consider the general exterior metric as
\begin{equation}\label{21}
ds^{2}=-e^{\eta_{+}(r)}dt^{2}+e^{\psi_{+}(r)}dr^{2}+r^{2}(d\theta^{2}+\sin^{2}\theta
d\phi^{2}).
\end{equation}
The continuity of the first fundamental form of matching conditions,
i.e., $[ds^{2}]_{\Sigma}=0$ ($\Sigma$ is hypersurface or star's
surface ($r=R$)) yields
\begin{eqnarray}\label{22}
\eta_{+}(R)=\eta_{-}(R),\quad 1-\frac{2M_{0}}{R}+\alpha
h^{\ast}(R)=e^{\psi_{+}(R)}.
\end{eqnarray}
Similarly, the continuity of second fundamental form
($[G_{\rho\gamma}t^{\gamma}]_{\Sigma}=0$, where $t^{\gamma}$ is a
unit four-vector in radial direction) leads to
\begin{eqnarray}\label{23}
P_{R}+\alpha
(\Theta_{1}^{1}(R))^{-}+(T_{1}^{1(\mathcal{G})}(R))^{-}=\alpha
(\Theta_{1}^{1}(R))^{+}+(T_{1}^{1(\mathcal{G})}(R))^{+}.
\end{eqnarray}

Using the matching condition (\ref{22}), we obtain
$(T_{1}^{1(\mathcal{G})}(R))^{-}=(T_{1}^{1(\mathcal{G})}(R))^{+}$
which implies that
\begin{eqnarray}\label{24}
P_{R}+\alpha (\Theta_{1}^{1}(R))^{-}=\alpha (\Theta_{1}^{1}(R))^{+},
\end{eqnarray}
yielding
\begin{eqnarray}\label{25}
P_{R}+\frac{\alpha h^{\ast}}{\kappa}
(\frac{\eta'}{R}+\frac{1}{R^{2}})=\frac{\alpha
f^{\ast}}{\kappa R^{2}}(\frac{R}{R-2M}),
\end{eqnarray}
where $f^{\ast}=0$ is the outer radial geometric deformation for
Schwarzschild metric given as
\begin{equation}\label{26}
ds^{2}=-(1-\frac{2m}{r})dt^{2}+(1-\frac{2m}{r}+\alpha
h^{\ast})^{-1}dr^{2}+r^{2}(d\theta^{2}+\sin^{2}\theta d\phi^{2}).
\end{equation}
The necessary and sufficient conditions for smooth matching of MGD
interior and exterior Schwarzschild metrics (filled by the field of
the source $\Theta_{\rho\gamma}$) are given by the constraints
Eqs.(\ref{22})-(\ref{25}). If the exterior geometric metric is taken
as the standard Schwarzschild metric ($f^{\ast}=0$) then
\begin{eqnarray}\label{27}
\bar{P}_{R}=P_{R}+\frac{\alpha h^{\ast}}{\kappa}
(\frac{\eta'}{R}+\frac{1}{R^{2}})=0.
\end{eqnarray}
In the following, we take a known isotropic spherically symmetric
solution for our systematic analysis.

\section{Interior Solutions}

In order to obtain anisotropic solution using MGD decoupling, it is
important to find out perfect fluid spherically symmetric solution.
In particular, we choose Krori and Barua solution for physical
relevance as \cite{24}
\begin{eqnarray}\label{a}
e^{\eta}&=&e^{Br^{2}+C},\\\label{a1}
e^{\chi}&=&\xi^{-1}(r)=e^{Ar^{2}},\\\label{a2}
\rho&=&\frac{e^{-Ar^{2}}}{\kappa
r^{2}}(e^{Ar^{2}}+2Ar^{2}-1)+T_{0}^{0(\mathcal{G})},\\\label{a3}
P&=&\frac{e^{-Ar^{2}}}{\kappa
r^{2}}(-e^{Ar^{2}}+2Br^{2}+1)-T_{1}^{1(\mathcal{G})},
\end{eqnarray}
where $A$, $B$ and $C$ are constants that can be derived through
matching conditions. The rationale for the aforementioned solution
is its singularity-free feature which satisfies physical conditions
inside the spherical distribution. For exterior geometric
configuration as Schwarzschild metric, the junction condition yields
\begin{eqnarray}\label{b}
A&=&-\frac{1}{R^{2}}\ln (1-\frac{2M_{0}}{R}),\quad
B=\frac{M_{0}}{R^{2}(R-2M_{0})},\\\label{b1} C&=&\ln
(\frac{R-2M_{0}}{R})-\frac{M_{0}}{R-2M_{0}},
\end{eqnarray}
with compactness $\frac{M_{0}}{R}<\frac{4}{9}$ ($M_{0}$ is the total
mass). The above expressions through the matching conditions ensure
the continuity of the interior and exterior regions at the boundary
of the star which definitely will vary in the presence of source
$\Theta_{\rho\gamma}$.

Now, we evaluate anisotropic solution, i.e., $\alpha\neq0$ in the
interior spherical distribution. The temporal and radial metric
contents are given by Eqs.(\ref{11}) and (\ref{a}) whereas the
geometric deformation and source are connected through
Eqs.(\ref{15})-(\ref{17}). For this purpose, various choices can be
considered such as the equation of state, some particular forms of
density as well as pressure or some physically motivated
restrictions on $h^{\ast}$ \cite{16,17,24a}. In any case, we need to remain
concerned with physical acceptability of the solution. In the
following, we address this problem by taking some conditions to
generate physically acceptable interior solutions.

\subsection{Solution-I}

Herein, we apply the constraint on source component $\Theta_{1}^{1}$
and solve the field equations for deformation function $h^{\ast}$
and source $\Theta_{\rho\gamma}$. One can observe that the exterior
geometry of Schwarzschild metric is compatible with interior matter
configuration as long as
$P+T_{1}^{1(\mathcal{G})}\sim\alpha(\Theta_{1}^{1}(R))_{-}$. The
simplest choice which satisfies this crucial requirement is
\begin{equation}\label{28}
\alpha\Theta_{1}^{1}=P+T_{1}^{1(\mathcal{G})}\quad\Rightarrow\quad
h^{\ast}=\xi-\frac{1}{1+r\eta'},
\end{equation}
where we have used Eqs.(\ref{13}) and (\ref{16}). Equation
(\ref{28}) mimics the radial metric component as
\begin{equation}\label{29}
e^{-\psi}=(1+\alpha)\xi-\frac{\alpha}{1+2Br^{2}}.
\end{equation}
The interior geometric component in Eqs.(\ref{a1}) and (\ref{29})
represent minimally deformed Krori and Barua solution by generic
anisotropic source $\Theta_{\rho\gamma}$. In the limit
$\alpha\rightarrow0$, Eq.(\ref{29}) yields standard isotropic
spherical solutions ((\ref{a})-(\ref{a3})).

The continuity of first fundamental form of junction conditions
gives
\begin{equation}\label{30}
Br^{2}+C=\ln(1-\frac{2M}{R}),
\end{equation}
\begin{equation}\label{31}
1-\frac{2M}{R}=(1+\alpha)\xi-\frac{\alpha}{1+2BR^{2}},
\end{equation}
whereas the continuity of second fundamental form
($P(R)+T_{1}^{1(\mathcal{G})}+\alpha(\Theta_{1}^{1}(R))_{-}=0$)
reads
\begin{equation}\label{32}
P(R)+T_{1}^{1(\mathcal{G})}=0\Rightarrow
A=\frac{\ln(1+2BR^{2})}{R^{2}},
\end{equation}
where expression in (\ref{28}) has been utilized. Using
Eq.(\ref{31}), the Schwarzschild mass is obtained as
\begin{equation}\label{33}
\frac{2M}{R}=\frac{2(1+\alpha)M_{0}}{R}+\frac{\alpha}{1+2BR^{2}}-\alpha.
\end{equation}
Using this in Eq.(\ref{30}), we have
\begin{equation}\label{34}
BR^{2}+C=\ln((1+\alpha)(1-\frac{2M_{0}}{R})-\frac{\alpha}{1+2BR^{2}}),
\end{equation}
where the constant $C$ is expressed in terms of $B$. Equations
(\ref{32})-(\ref{34}) give the necessary and sufficient conditions
for smooth matching of interior and exterior spacetimes on the
surface of star. Using the mimic constraint (\ref{28}), the
anisotropic solution ($\bar{\rho}, \bar{P}_{r}, \bar{P}_{t}$) is
given by
\begin{eqnarray}\nonumber
\bar{\rho}&=&\frac{1}{\kappa(r+2Br^{3})^{2}}[(1+2(1+\alpha)
Ar^{2}e^{-Ar^{2}}+(1-\alpha)e^{-Ar^{2}}+\kappa
r^{2}T_{0}^{0(\mathcal{G})})\\\label{d3}&\times&(1+4Br^{2}(1+Br^{2}))+\alpha(1-2Br^{2})],\\\label{d3a}
\bar{P}_{r}&=&\frac{1}{\kappa
r^{2}}(1+\alpha)((2Br^{2}+1)e^{-Ar^{2}}-1+\kappa
r^{2}T_{1}^{1(\mathcal{G})}),\\\nonumber
\bar{P}_{t}&=&\frac{1}{\kappa(r+2Br^{3})^{2}}[\alpha
Ar^{2}e^{-Ar^{2}}(1+5Br^{2}+8B^{2}r^{4}+4b^{3}r^{6})+Br^{2}e^{-Ar^{2}}\\\nonumber&\times&
(-2(3+\alpha)+3Br^{2}(3\alpha-4)-4B^{2}r^{4}(2+3\alpha)-4\alpha
B^{3}r^{6})+\alpha B^{2}r^{4}\\\label{d4}&\times&
(3+2Br^{2})-e^{-Ar^{2}}+(1+\kappa
r^{2}T_{1}^{1(\mathcal{G})})(1+4Br^{2}+4B^{2}r^{4})],\\\nonumber
\bar{\Delta}&=&\bar{P}_{t}-\bar{P}_{r}=\frac{1}{\kappa(r+2Br^{3})^{2}}
[1+e^{-Ar^{2}}(-1-Ar^{2}(1+5Br^{2}+8B^{2}r^{4}\\\nonumber&+&4B^{3}r^{6})+Br^{2}(-4-3Br^{2}
+4B^{2}r^{4}+4B^{3}r^{6}))+B^{2}r^{4}+4Br^{2}\\\label{d5}&-&2B^{3}r^{6}+\kappa
T_{1}^{1(\mathcal{G})}r^{2}(1+4Br^{2}+4B^{2}r^{4})].
\end{eqnarray}

\subsection{Solution-II}

In this case, we take another choice of mimic constraint on density
for physically acceptable solution. This constraint
($\alpha\Theta_{0}^{0}\sim\rho-T_{0}^{0(\mathcal{G})}$) implies that
\begin{equation}\label{35}
h^{\ast}=\frac{c_{1}}{r}+e^{-Ar^{2}}-1,
\end{equation}
where $c_{1}$ is the integration constant. By adopting the same
methodology as prescribed in solution-I, we obtain the junction
conditions as
\begin{eqnarray}\label{36}
&&\alpha R(e^{-AR^{2}}-1)+\alpha c_{1}+2(M-M_{0})=0,\\\label{37}
BR^{2}&+&C=\ln(1-\frac{2M_{0}}{R}+\frac{\alpha c_{1}}{R}+
\alpha (e^{-AR^{2}}-1),
\end{eqnarray}
The anisotropic solution satisfying the above matching condition
becomes
\begin{eqnarray}\label{39}
\bar{\rho}&=&(1-\alpha)[\frac{e^{-Ar^{2}}}{\kappa
r^{2}}(e^{Ar^{2}}+2Ar^{2}-1)+T_{0}^{0(\mathcal{G})}],\\\nonumber
\bar{P}_{r}&=&\frac{1}{\kappa
r^{3}}[2Be^{-Ar^{2}}r^{3}(\alpha+1)+2\alpha Br^{2}(c_{1}-r)+re^{-Ar^{2}}(\alpha+1)-
(1+\alpha)r\\\label{40}&-&T_{1}^{1(\mathcal{G})}\kappa r^{3},\\\nonumber
\bar{P}_{t}&=&-\frac{1}{2\kappa
r^{3}}[2Be^{-Ar^{2}}\alpha r^{5}(A-B)+2\alpha B^{2}r^{4}(r-c_{1})-4Be^{-Ar^{2}}r^{3}(1+\alpha)\\\label{41}&+&4B\alpha r^{2}(4r-3c_{1})+2re^{-Ar^{2}}(Ar^{2}-1)+2r+\alpha c_{1}+2\kappa r^{3}T_{1}^{1(\mathcal{G})}]
\end{eqnarray}
with anisotropic parameter
\begin{eqnarray}\nonumber
\bar{\Delta}=\bar{P}_{t}-\bar{P}_{r}&=&-\frac{\alpha}{2\kappa
r^{3}}[2B e^{-Ar^{2}}r^{5}(A-B)+2B^{2}r^{4}(r-c_{1})+2re^{-Ar^{2}}
\\\label{42}&\times&(Ar^{2}+1)+Bc_{1}r^{2}+3c_{1}+2r].
\end{eqnarray}

\subsection{Graphical Analysis of Some Specific Solutions}

In this section, we analyze anisotropic matter distribution
prescribed by solutions I and II for specific form of generic
function as
\begin{equation}\label{43}
f(\mathcal{G})=\beta \mathcal{G}^{n},
\end{equation}
where $\beta$ is constant and $n>0$ \cite{25}. In order to examine
the solution-I graphically, we use $n=2$ and $\beta=0.25$ while the
constant $A$ is taken from Eq.(\ref{32}). The free parameters $B$
and $C$ are fixed through matching conditions for isotropic
distribution given in Eqs.(\ref{b}) and (\ref{b1}). For compact
star, the energy density as well as radial pressure must be
positive, finite and maximum at its core, i.e., it must obey
monotonically decreasing behavior as radial component $r$ increases.
The plot of effective energy density $\bar{\rho}$ is shown in Figure
\textbf{1} (left plot, row-1). It is observed that density gives the
maximum value at the interior of star and decreases gradually as $r$
increases. It is also found that the density increases with increase
in $\beta$ which presents more dense spherical structure while it
decreases with increase in decoupling constant $\alpha$. This
demonstrates that the model becomes less dense in the presence of
coupling parameter.
\begin{figure}\center
\epsfig{file=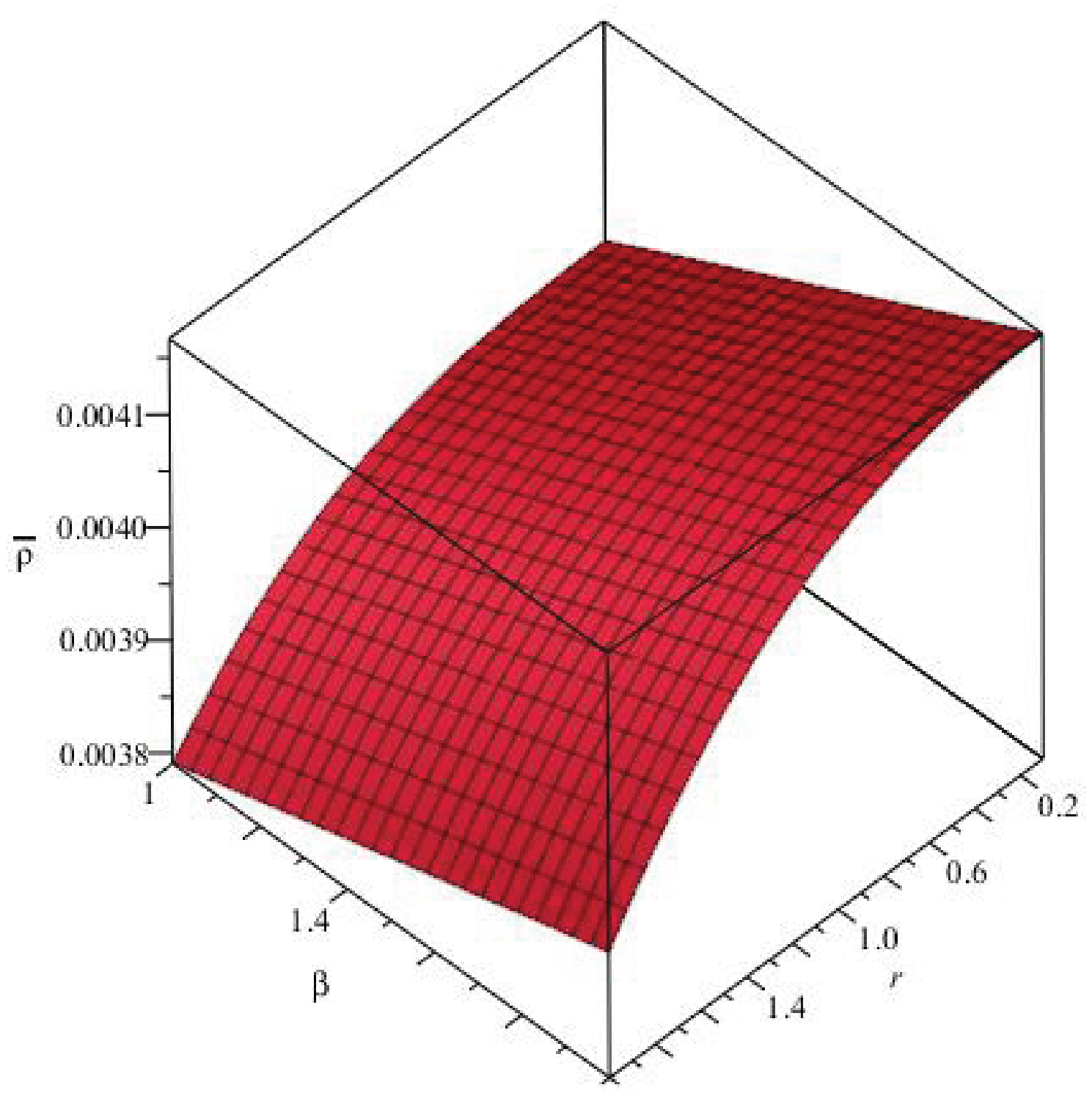,width=0.45\linewidth}\epsfig{file=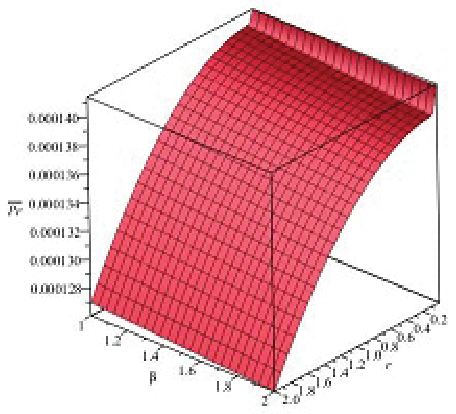,width=0.45\linewidth}
\epsfig{file=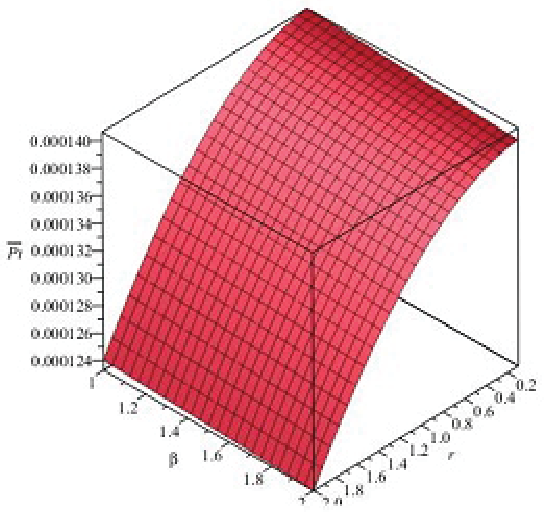,width=0.45\linewidth}\epsfig{file=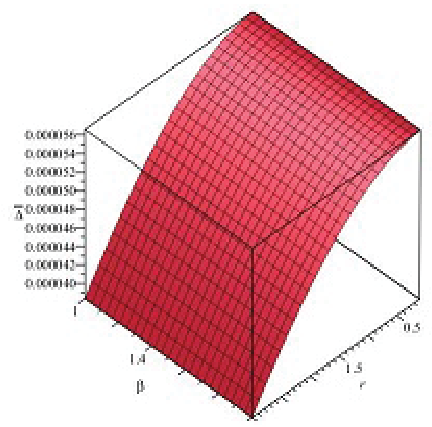,width=0.45\linewidth}
\caption{Plots of $\bar{\rho}$ (left plot, row-1), $\bar{P}_{r}$
(right plot, row-1), $\bar{P}_{t}$ (left plot, row-2) and
$\bar{\Delta}$ (right plot, row-2) against $r$ using
$M_{0}=1M_{\odot}$ and $R=(0.2)^{-1}M_{\odot}$ for solution-I.}
\end{figure}

The radial ($\bar{P}_{r}$) and tangential pressures ($\bar{P}_{t}$)
show the same pattern against radius of stellar object while
tangential pressure decreases with increase in $\beta$ as compared
to radial (remains the same). It is seen that the radial pressure
decreases when $\alpha$ increases as compared to the inverse
behavior of tangential pressure for $\alpha$. The anisotropy
parameter $\bar{\Delta}$ gives necessary information about
anisotropy of the fluid configuration. For
$\bar{P}_{t}>\bar{P}_{r}$, the anisotropy parameter remains positive
which shows that it is outward directed while for
$\bar{P}_{t}<\bar{P}_{r}$, it corresponds to inward directed. In our
case, we measure anisotropy outward directed scenario (Figure
\textbf{1}, right plot, row-2). Moreover, the generic anisotropy
remains the same for $\beta$ while increases with increase in
$\alpha$. It is crucial to check the viability of the resulting
solutions. For this purpose, we investigate the energy conditions
which describe physically realistic matter configuration. The
corresponding energy conditions are defined as
\begin{equation}\nonumber
\bar{\rho}\geq0,\quad\bar{\rho}+\bar{P}_{r}\geq0,\quad
\bar{\rho}+\bar{P}_{t}\geq0,\quad\bar{\rho}-\bar{P}_{r}\geq0,
\end{equation}
\begin{equation}\nonumber
\bar{\rho}-\bar{P}_{t}\geq0,\quad
\bar{\rho}+\bar{P}_{r}+2\bar{P}_{t}\geq0.
\end{equation}
\begin{figure}\center
\epsfig{file=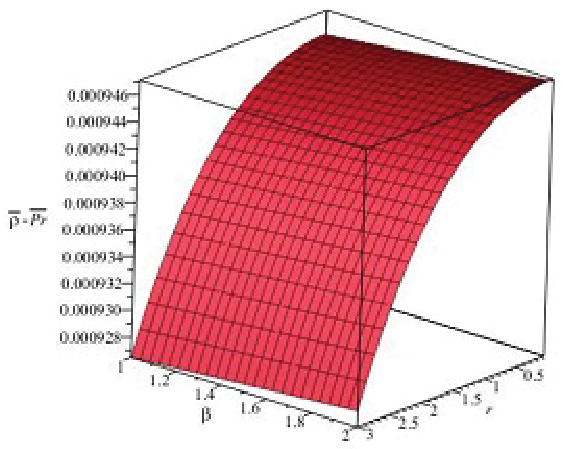,width=0.45\linewidth}\epsfig{file=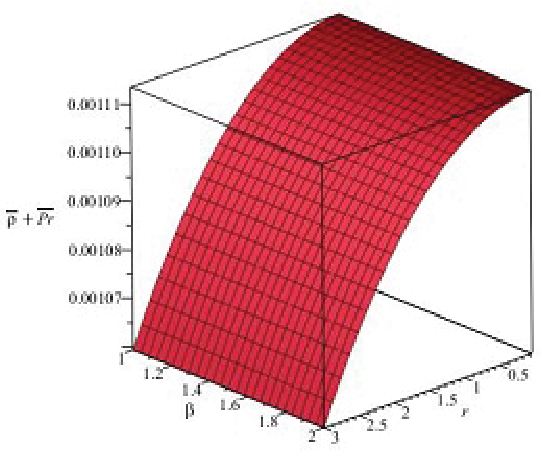,width=0.45\linewidth}
\epsfig{file=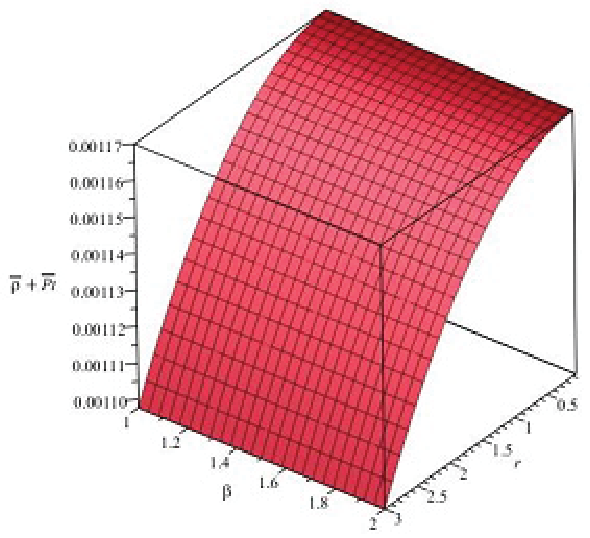,width=0.45\linewidth}\epsfig{file=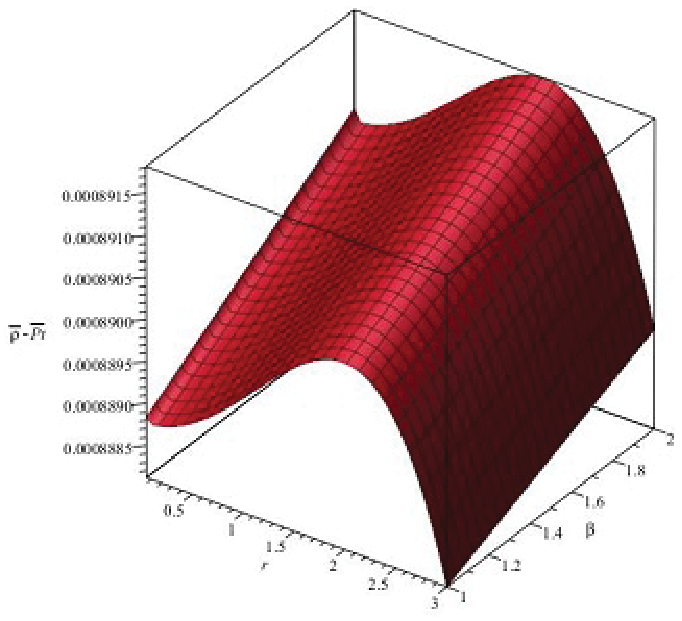,width=0.45\linewidth}
\epsfig{file=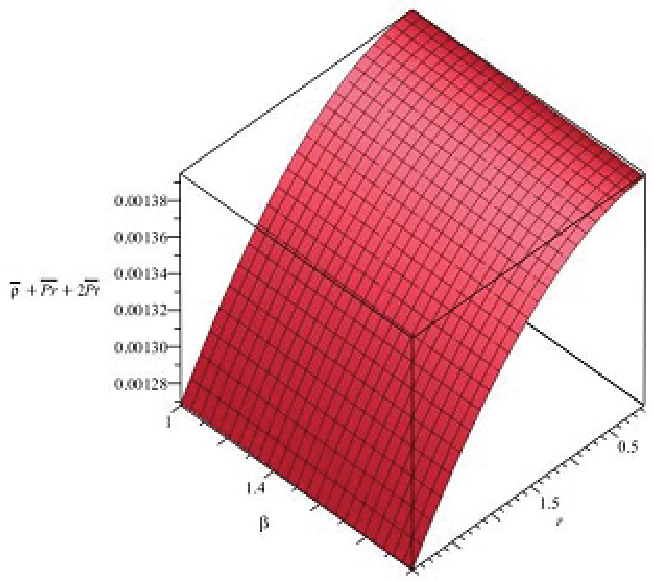,width=0.45\linewidth} \caption{Plots of energy
conditions against $r$ and $\beta$ using $M_{0}=1M_{\odot}$ and
$R=(0.2)^{-1}M_{\odot}$ for solution-I.}
\end{figure}
\begin{figure}\center
\epsfig{file=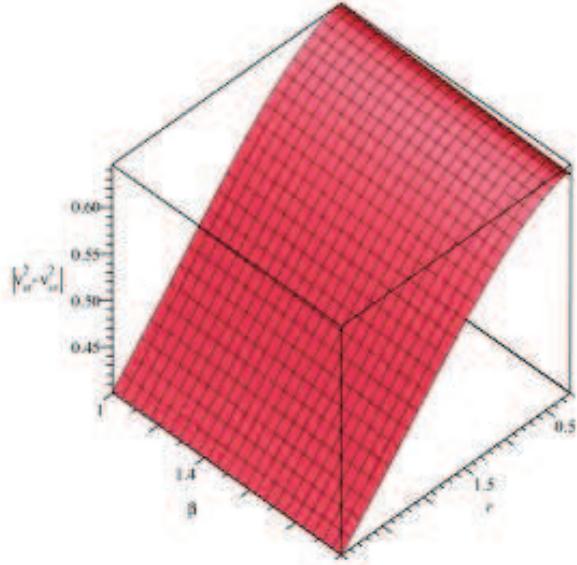,width=0.75\linewidth}\caption{Plots of
$|\nu_{st}^{2}-\nu_{sr}^{2}|$ against $r$ and $\beta$ using
$M_{0}=1M_{\odot}$ and $R=(0.2)^{-1}M_{\odot}$ for solution-I.}
\end{figure}
\begin{figure}\center
\epsfig{file=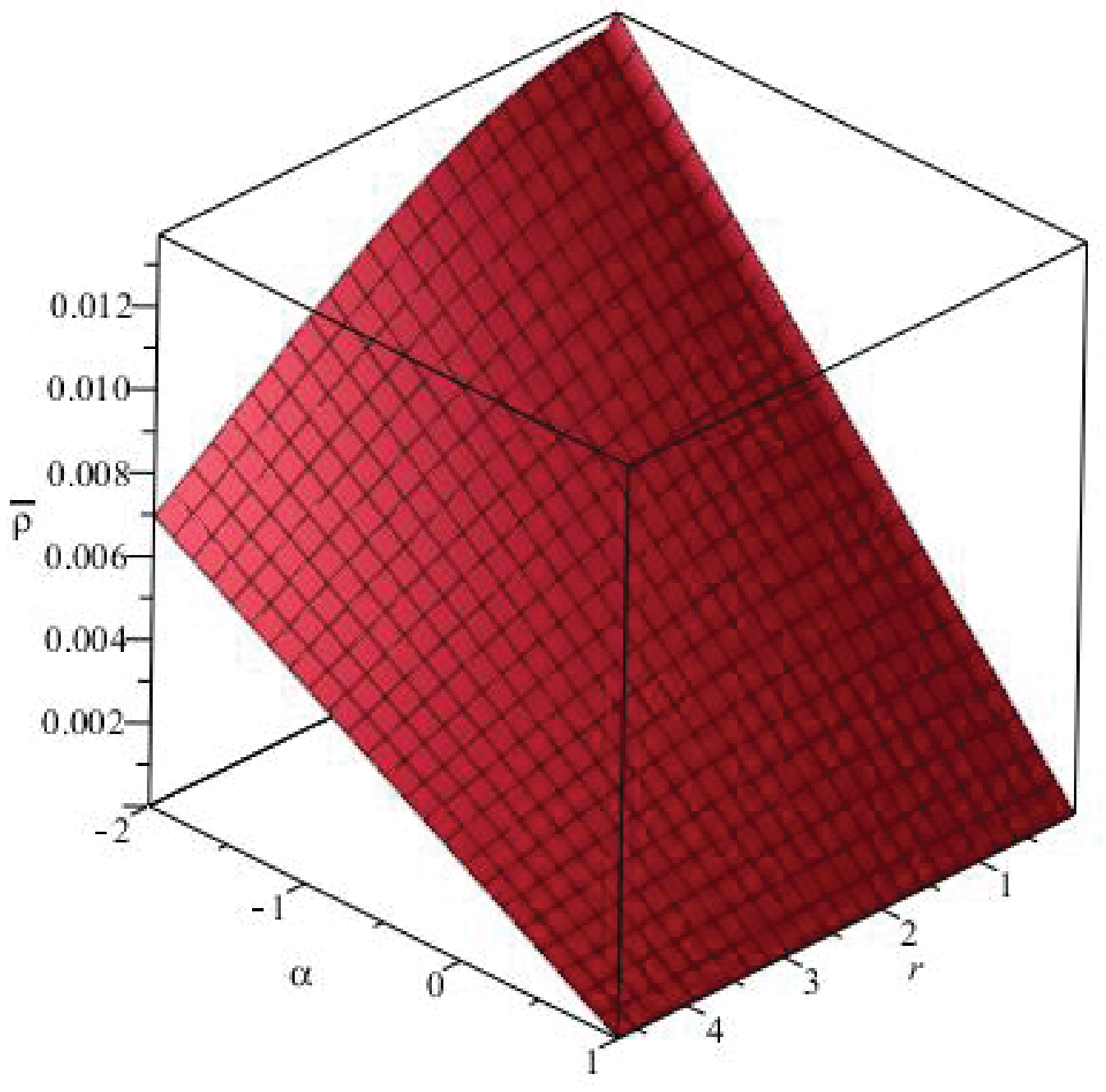,width=0.45\linewidth}\epsfig{file=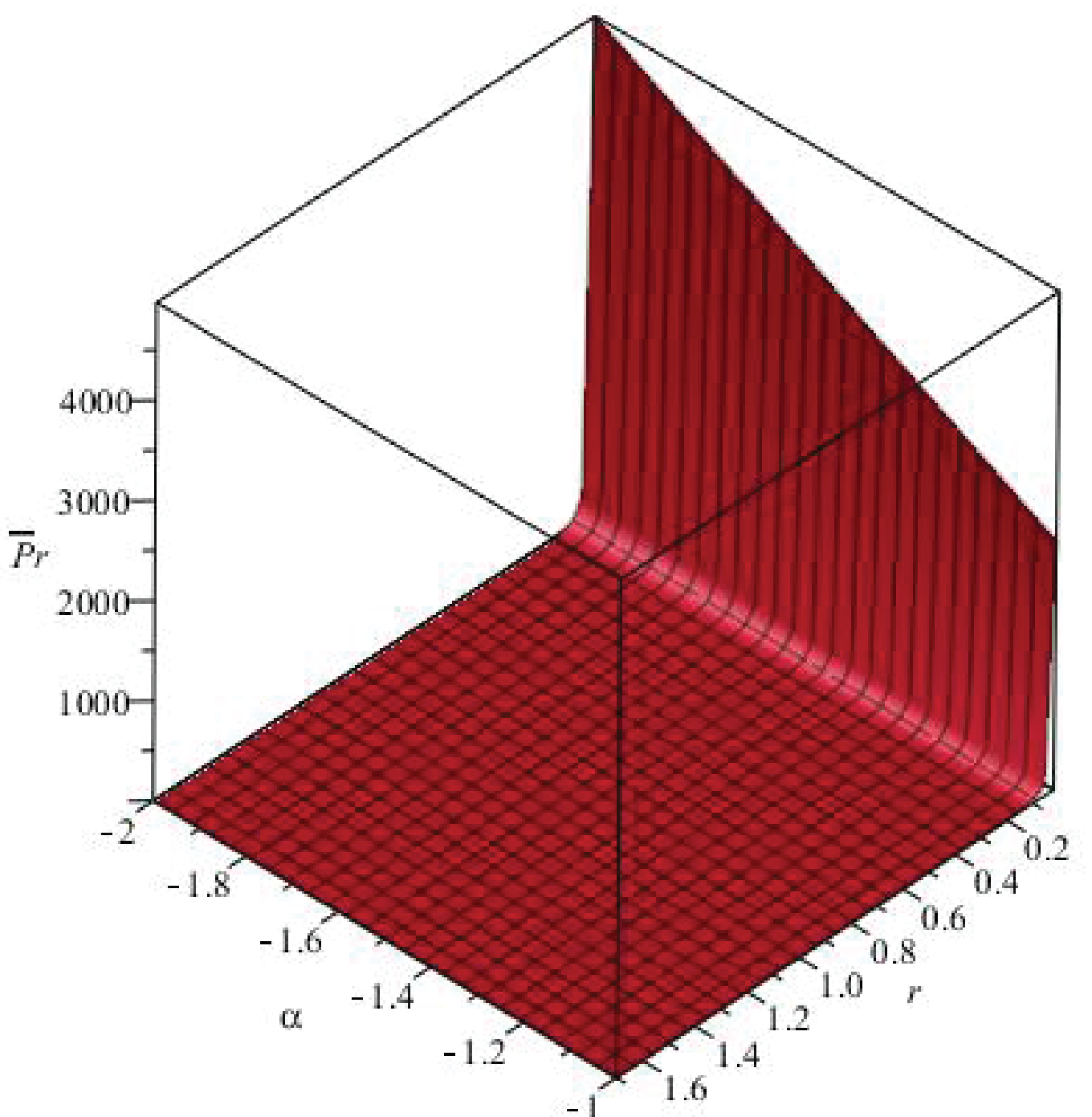,width=0.45\linewidth}
\epsfig{file=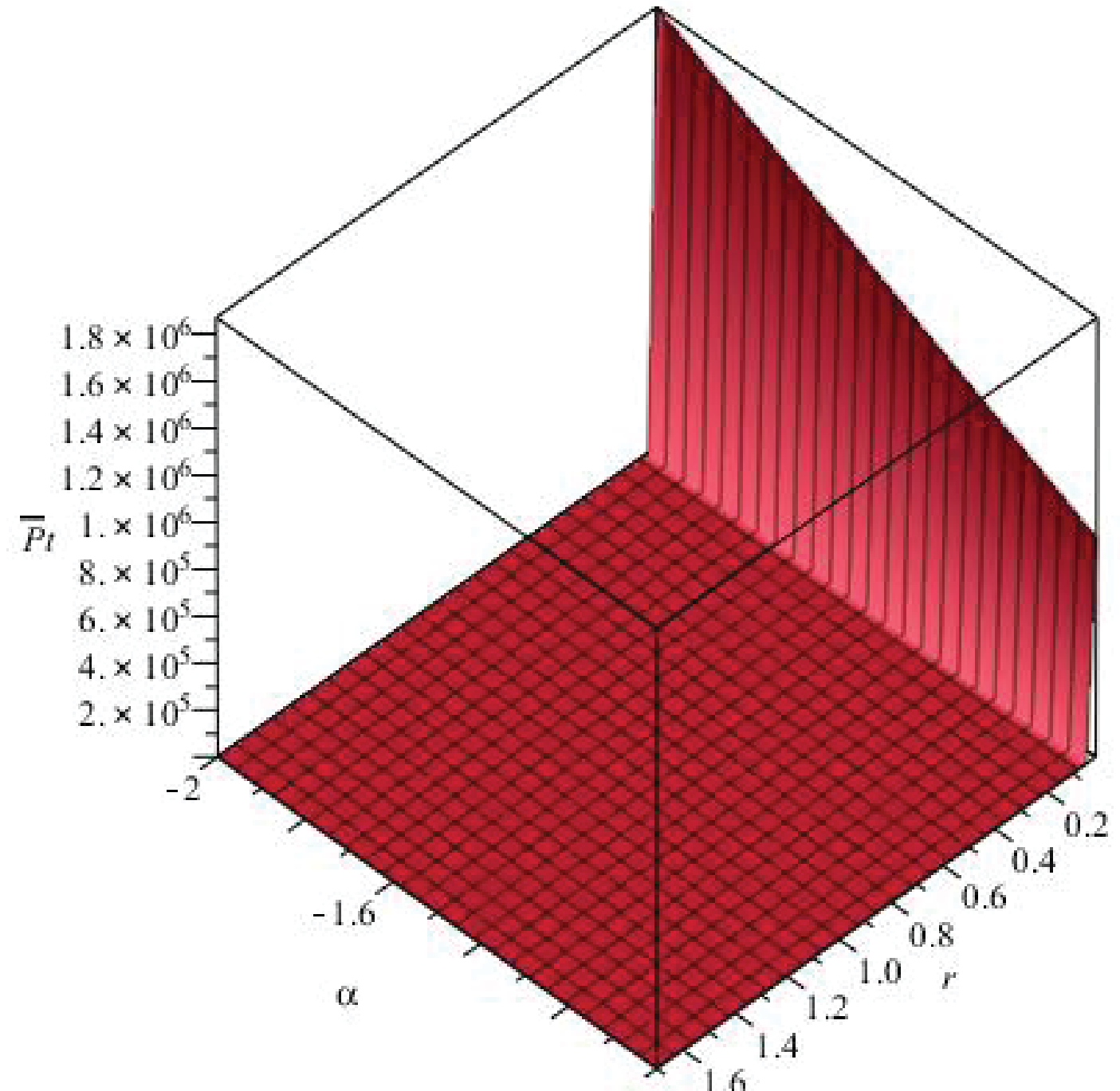,width=0.45\linewidth}\epsfig{file=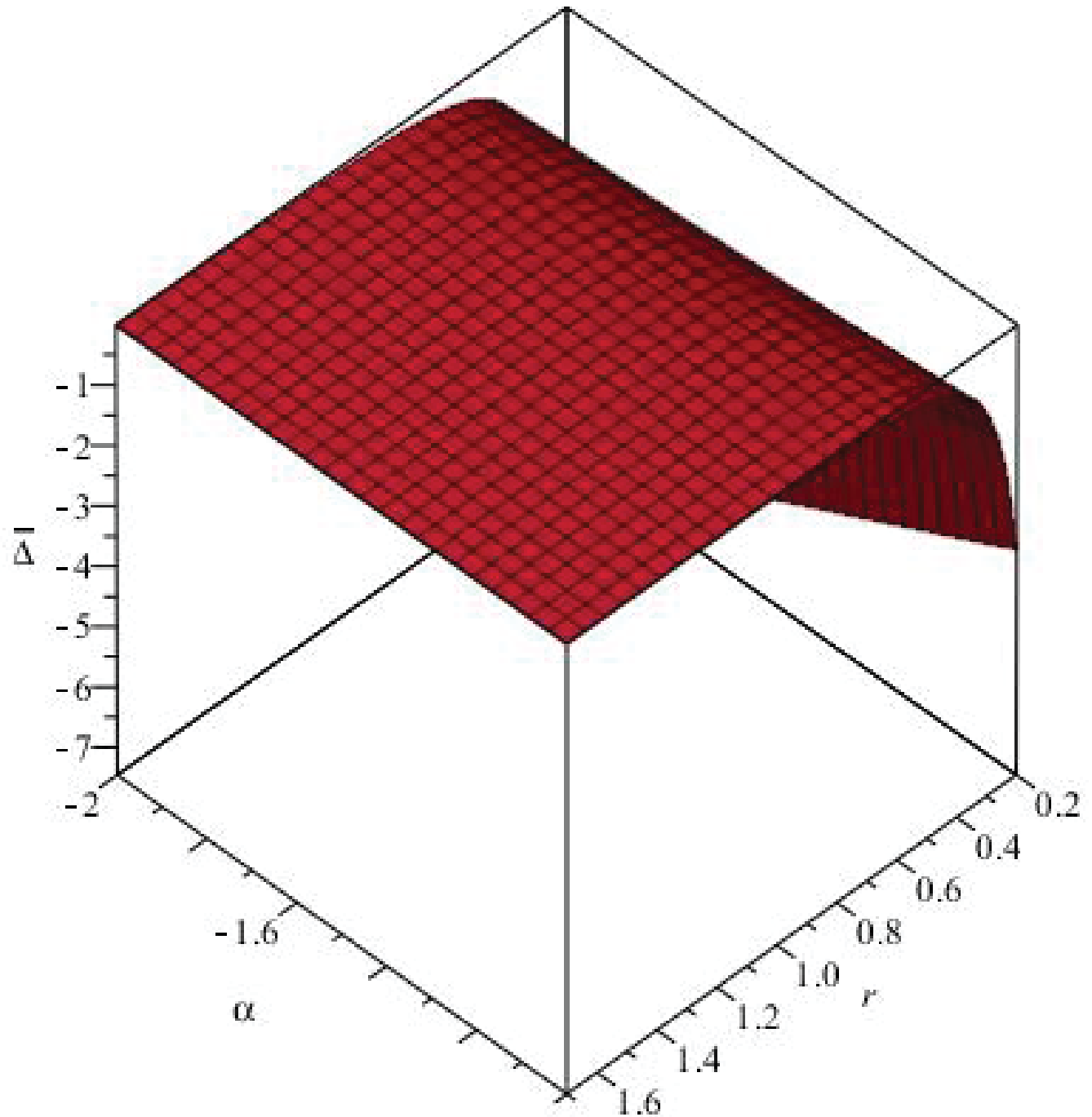,width=0.45\linewidth}
\caption{Plots of $\bar{\rho}$ (left plot, row-1), $\bar{P}_{r}$
(right plot, row-1), $\bar{P}_{t}$ (left plot, row-2) and
$\bar{\Delta}$ (right plot, row-2) against $r$ and $\alpha$ using
$M_{0}=1M_{\odot}$ and $R=(0.3)^{-1}M_{\odot}$ for solution-II.}
\end{figure}
\begin{figure}\center
\epsfig{file=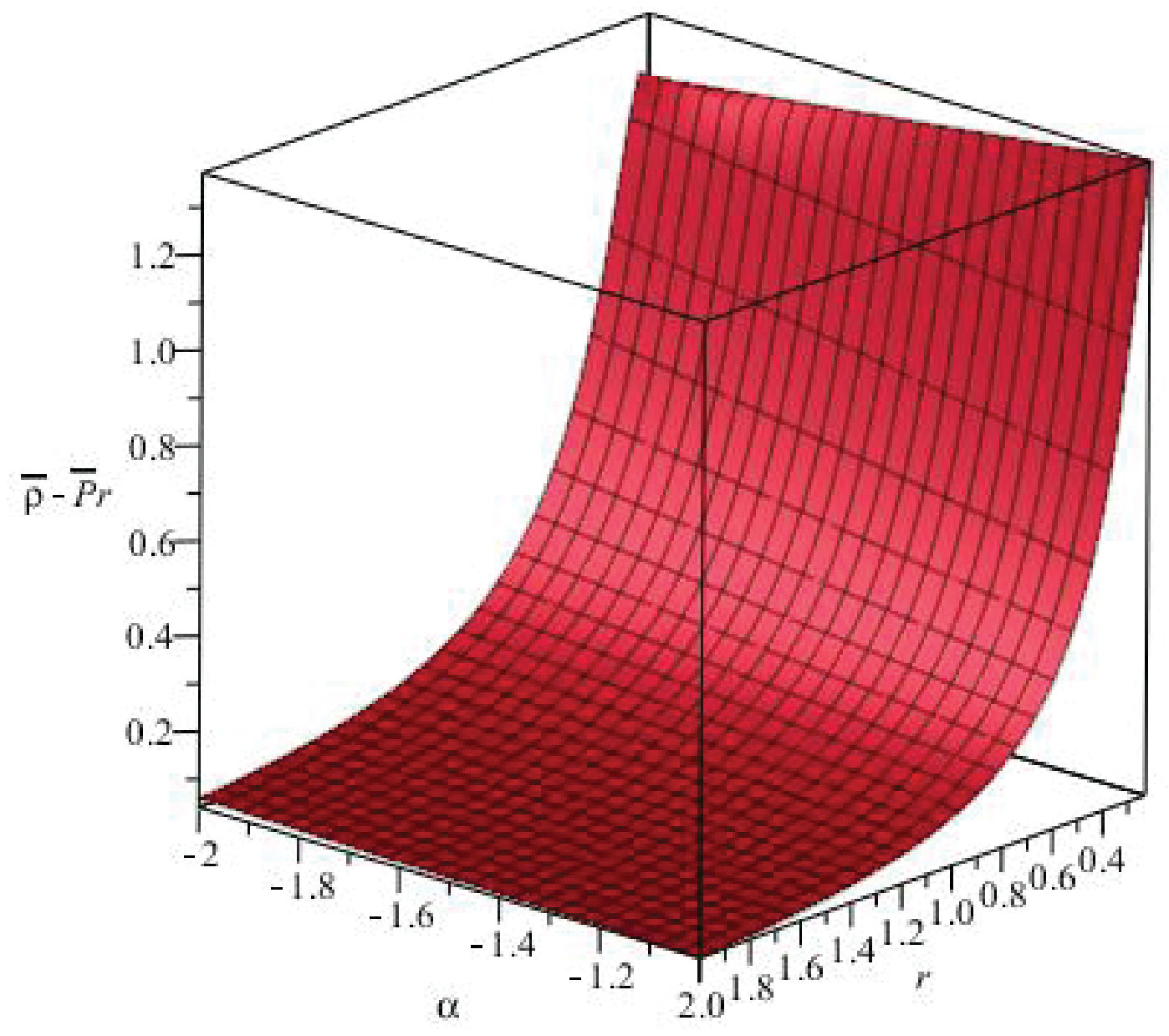,width=0.45\linewidth}\epsfig{file=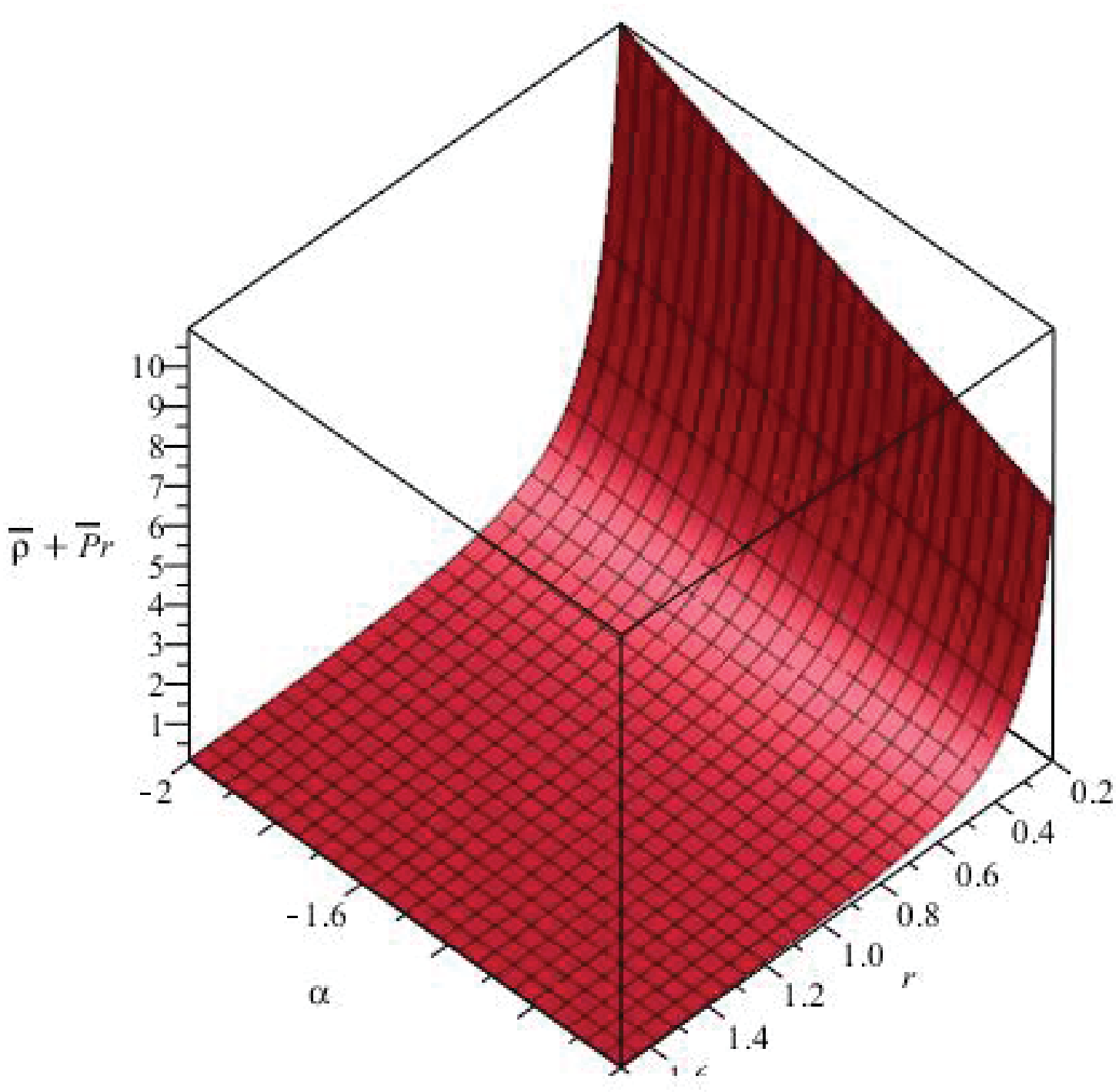,width=0.45\linewidth}
\epsfig{file=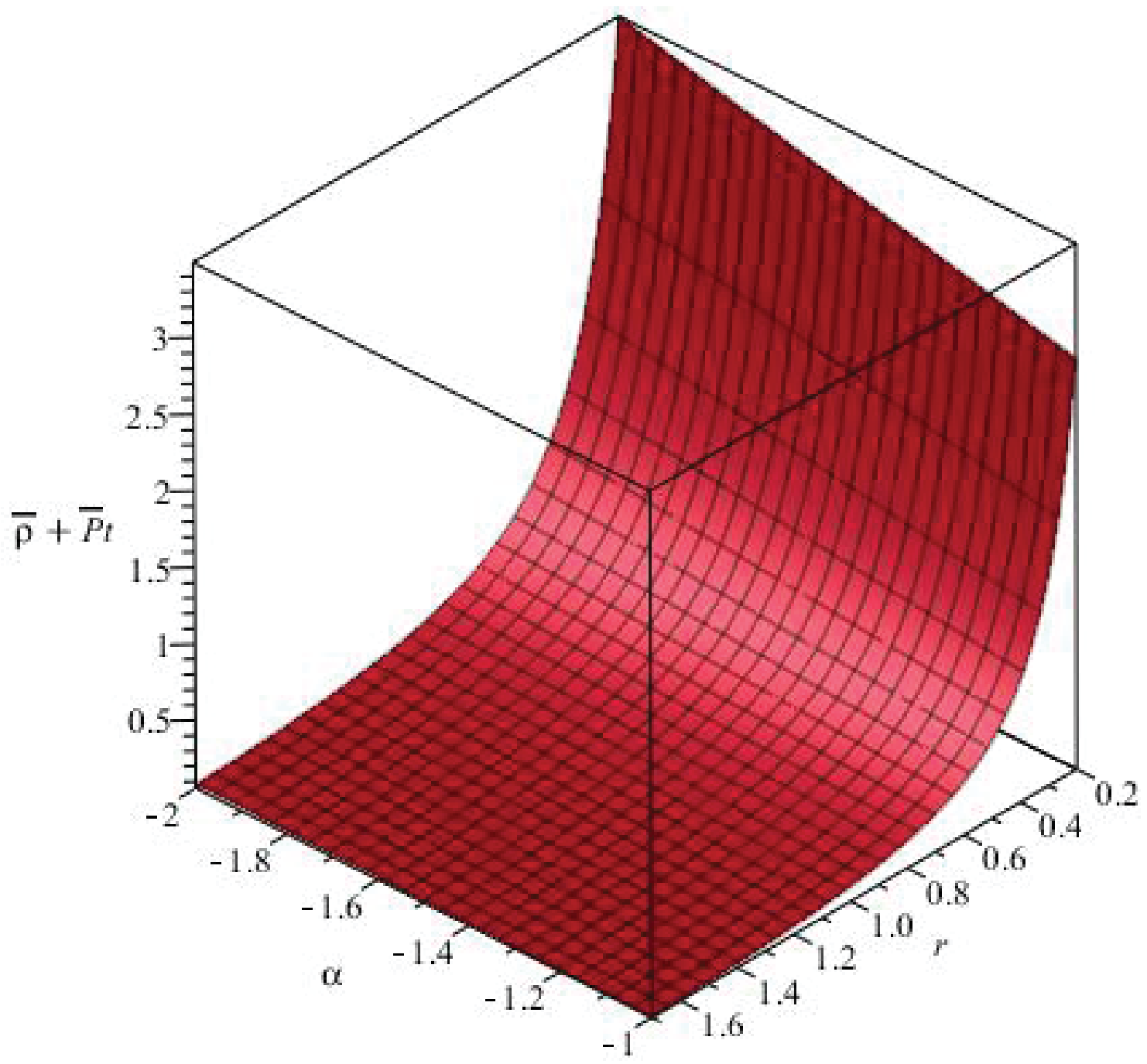,width=0.45\linewidth}\epsfig{file=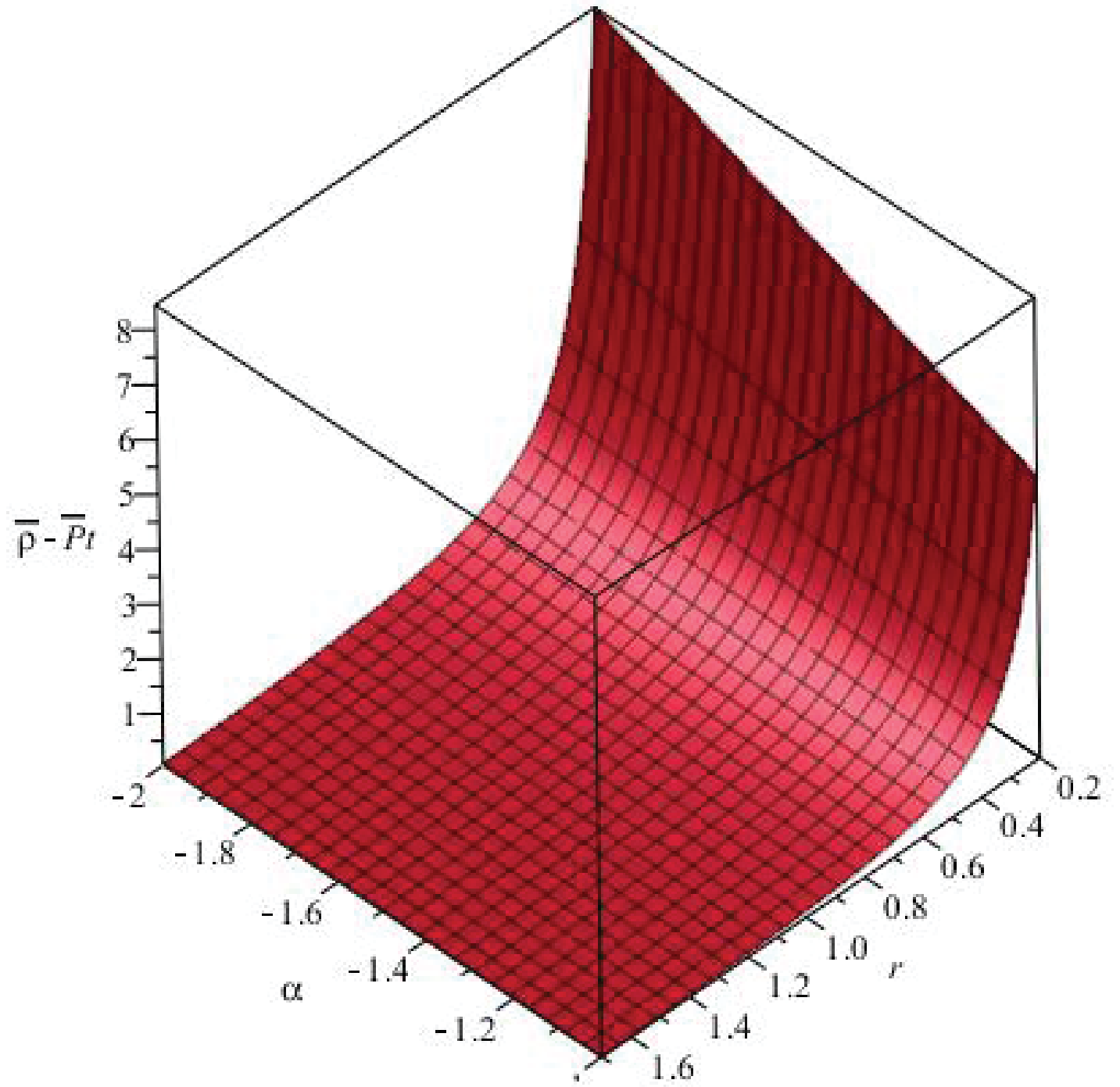,width=0.45\linewidth}
\epsfig{file=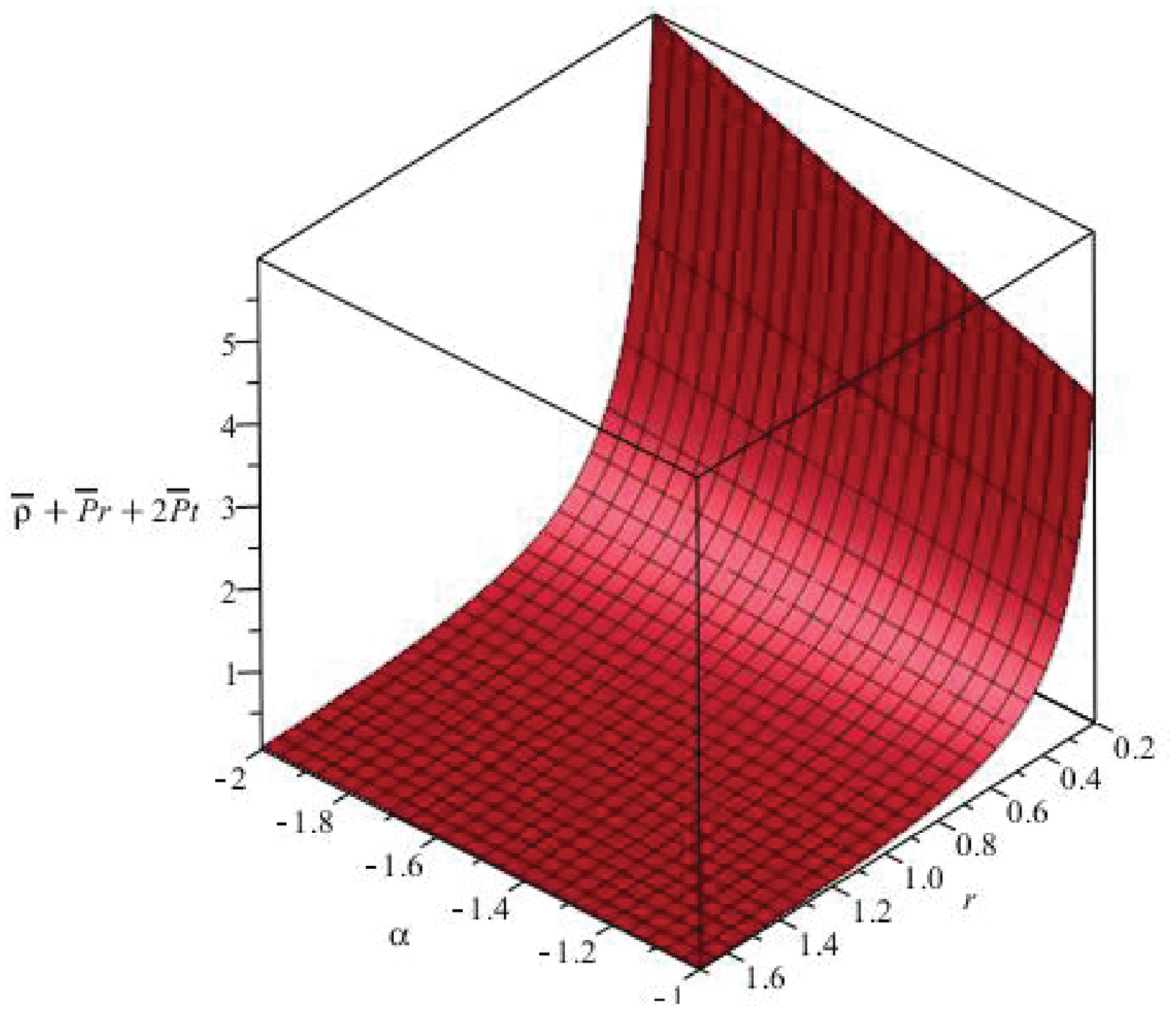,width=0.45\linewidth} \caption{Plots of energy
conditions against $r$ and $\alpha$ using $M_{0}=1M_{\odot}$ and
$R=(0.3)^{-1}M_{\odot}$ for solution-II.}
\end{figure}
\begin{figure}\center
\epsfig{file=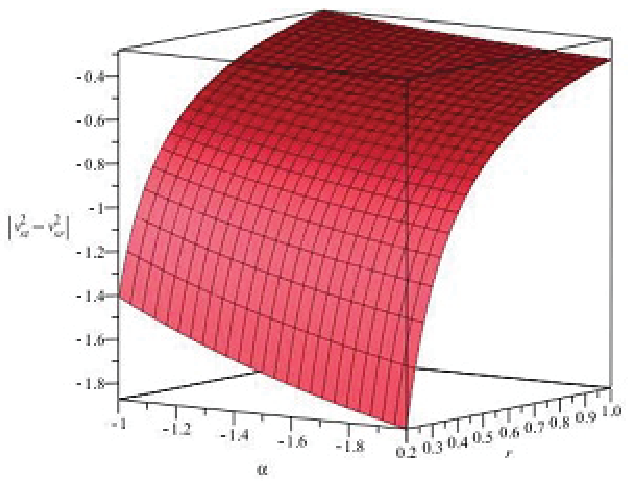,width=0.65\linewidth}\caption{Plots of
$|\nu_{st}^{2}-\nu_{sr}^{2}|$ against $r$ and $\alpha$ using
$M_{0}=1M_{\odot}$ and $R=(0.3)^{-1}M_{\odot}$ for solution-II.}
\end{figure}

These conditions are shown graphically for the derived anisotropic
solution in Figure \textbf{2} which indicates the validity of all
energy condition and hence viability of the resulting anisotropic
solution. For the stability analysis, we plot the squared speed of
sound as shown in Figure \textbf{3}. For stability of the solution,
the condition $0<|\nu_{st}^{2}-\nu_{sr}^{2}|<1$, must be satisfied.
Figure \textbf{3} shows $|\nu_{st}^{2}-\nu_{sr}^{2}|\leq1$ for small
$\alpha$ but it is violated for its large value.

For analyzing physical characteristics of solution-II, we choose
constants $A$, $B$, $C$ from the matching conditions (\ref{b}) and
(\ref{36})-(\ref{37}) whereas free parameters set as $c_{1}=-0.5$,
$\beta=1.2$ and $n=3$, respectively. It is observed that the behavior of
$\bar{\rho}$, $\bar{P}_{r}$ and $\bar{P}_{t}$ against $r$ is
consistent with solution-I whereas both the radial and tangential pressure
decreases linearly with increase in parameter $\alpha$ (Figure \textbf{4}).
The plot of these matter contents shows that as $r\rightarrow0$,
they attain their maximum and this, in fact, indicates the high
compactness of the core of the star validating that our model
under analysis is viable for the outer region of the core.
Moreover, anisotropy is greater at the surface of star than that of core which
shows opposite behavior as compared to solution-I. The plots of
energy conditions (Figure \textbf{5}) show that our solution
satisfies all energy bounds which suffices for the viable solution.
However, the stability condition is violated (Figure \textbf{6}).

\section{Concluding Remarks}

Recently, the minimal gravitational decoupling approach has
extensively been used to obtain exact solutions for interior
configuration of stellar distribution. In this work, we have used
MGD decoupling technique in $f(\mathcal{G})$ gravity to extend
interior isotropic spherical solution to anisotropic solution
contained in gravitational source. For this purpose, we have
introduced a new source in isotropic energy-momentum tensor
constituting the field equations for anisotropic matter
distribution. We have introduced minimal geometrical deformation in
metric functions (radial metric component only). It is found that
the corresponding field equations with source can be split into two
systems: one corresponds to standard field equations of
$f(\mathcal{G})$ gravity and other contains the source term with
deformed coefficients. These two systems express that there is a
purely gravitational interaction, without direct exchange of
energies.

We have studied junction condition for smooth matching of interior
and exterior geometries for the deformed Schwarzschild spacetime.
For anisotropic solution, we have taken a known solution then
extended it through the source added in perfect fluid geometry. We
have imposed constraints on the effective energy density and
effective pressure which constitute solutions I and II,
respectively. For physical acceptability, we have introduced
specific generic function and examined energy conditions as well as
squared speed of sound and anisotropy parameter. We have observed
that solution-I as well as solution-II are physically acceptable and
corresponds to stability of stellar object. Finally, we would like
to mention here that our results for the first solution are
consistent with those obtained in general relativity (GR)
\cite{16,17}. It is also worth mentioning here that modified
$f(\mathcal{G})$ gravity provides viable spherically symmetric
solutions (as the energy conditions for both solutions are
satisfied) due to inclusion of correction terms as compared to GR in
which the second type of solutions do not meet the energy bounds
\cite{16,17}.

\section*{Appendix A}

The GB corrections for the interior metric are
\begin{eqnarray}\nonumber
\kappa
T_{0}^{0(\mathcal{G})}&=&\frac{1}{2}f(\mathcal{G})+\frac{e^{-2\psi}}{2r^{3}}[\eta'^{2}(2r(9-\eta'\psi'
r^{2}+4r\psi'-6r\psi')+r^{3}(\eta'^{2}+\psi'^{2})\\\nonumber&+&4\eta'(4-r\psi'+r^{2}\psi'^{2}))+\exp(\psi)
(\eta'(-16+2r\psi'-2r\eta')-4r\eta'')\\\nonumber&+&4r\eta''(1-r^{2}\eta'\psi'+r^{2}(1+\eta'^{2})+4r\eta'
-2r\psi')]\frac{df(\mathcal{G})}{d\mathcal{G}}-[\mathcal{G}'(-8\\\nonumber&+&r^{3}\psi'\eta'^{2}-r^{3}\psi'\eta'
-2e^{\psi}r\psi'+4r^{2}\psi'\eta'+2r^{3}\psi'\eta''-4r^{2}\psi'^{2}+6r\psi'\\\nonumber&+&8e^{\psi})
+\mathcal{G}''(4r(e^{\psi}-1)+8r^{2}(\psi'-\eta')+3r^{3}(\eta'\psi'-\eta'^{2}-2\eta''))]
\frac{e^{-2\psi}}{r^{3}}\\\nonumber&\times&\frac{d^{2}f(\mathcal{G})}{d\mathcal{G}^{2}}+
\frac{2e^{-2\psi}}{r^{2}}[r^{2}(2\eta''-\eta'^{2}-\eta'\psi')+4r(\eta'-\psi')\\\nonumber&-&
2e^{-\psi}(1-e^{-\psi})]\mathcal{G}'^{2}\frac{d^{3}f(\mathcal{G})}{d\mathcal{G}^{3}},
\end{eqnarray}
\begin{eqnarray}\nonumber
\kappa
T_{1}^{1(\mathcal{G})}&=&\frac{1}{2}f(\mathcal{G})+\frac{e^{-2\psi}}{2r^{3}}
[4r\eta'^{2}(1-e^{\psi}+r^{2}\eta''-4r\psi')+4r^{2}\eta'\eta''(r(\eta'-\psi')
\\\nonumber&+&2)+\eta'^{2}(2r(1+2r\eta'-e^{\psi}-r^{2}\eta'\psi'-6r\psi'')
+r^{3}(\eta'-\psi'))\\\nonumber&+&2r\eta'\psi'(-7+e^{\psi}+4r\psi')
-16\psi'(1-\psi'-e^{\psi})]\frac{df(\mathcal{G})}{d\mathcal{G}}
\\\nonumber&+&\frac{e^{-2\psi}}{2r^{3}}[2r^{2}\eta''
(-8-\eta'r+e^{\eta-\psi}r\eta'-2r\psi')+\eta'(4r(-7+e^{\psi}\\\nonumber&+&3r\psi'
-e^{\eta-\psi}r\psi')-r^{3}\psi'^{2})]
\mathcal{G}'\frac{d^{2}f(\mathcal{G})}{d\mathcal{G}^{2}}
-\frac{e^{-2\psi}}{r^{2}}[r^{2}(2\eta''+\eta'^{2}-\eta'\psi')
\\\nonumber&+&4r(\eta'-\psi'-e^{\psi})+4]
\mathcal{G}'^{2}\frac{d^{3}f(\mathcal{G})}{d\mathcal{G}^{3}},
\end{eqnarray}
\begin{eqnarray}\nonumber
\kappa
T_{2}^{2(\mathcal{G})}&=&\frac{1}{2}f(\mathcal{G})+\frac{e^{-2\psi}}{r^{4}}
[r^{2}\eta'^{2}(5-e^{\psi}+r\eta'-2r\psi')+r\eta'(12-7r\psi'
\\\nonumber&+&re^{\psi}\psi'-12e^{\psi}+r^{2}\psi'^{2})+2r^{2}\eta''(1-e^{\psi}+r \eta'-r
\psi')-12r\psi'\\\nonumber&-&16e^{\psi}+8e^{2\psi}+12re^{\psi}\psi'+4r^{2}\psi'^{2}+8]
\frac{df(\mathcal{G})}{d\mathcal{G}}+\frac{e^{-2\psi}}{2r^{3}}[4r^{2}\eta''\mathcal{G}''
\\\nonumber&+&\mathcal{G}'(\eta'(4r^{2}
-r^{3}\psi'^{2}-8r^{2}\psi'+12r)+\eta'^{2}r^{2}(r\psi'+6)+2r^{2}\eta''(r\psi'-6)\\\nonumber&-&
8(e^{\psi}-1)-2r\psi'(r\psi'+6))]\frac{d^{2}f(\mathcal{G})}{d\mathcal{G}^{2}}
-2\frac{e^{-2\psi}}{r^{2}}\eta'r\mathcal{G}'^{2}\frac{d^{3}f(\mathcal{G})}{d\mathcal{G}^{3}}.
\end{eqnarray}

\end{document}